\documentclass[twocol]{ametsocV6.1}

\title{The Climatological Relationship Between United States Tornadoes and Extratropical Cyclones\\
{\color{red}NOT PUBLISHED. In peer review (under revision).}}

\authors{Lauren A. Kiefer,\aff{a}\correspondingauthor{Lauren A. Kiefer, kieferl@purdue.edu}
Daniel R. Chavas,\aff{a} 
Michelle J. Gore,\aff{b} 
Daniel T. Dawson II,\aff{a}
}

\affiliation{\aff{a}{Department of Earth, Planetary, and Atmospheric Sciences, Purdue University, West Lafayette, Indiana, USA}\\
\aff{b}{Department of Meteorology and Atmospheric Science, The Pennsylvania State University, University Park, Pennsylvania, USA,}
}

\abstract{Tornadoes have caused billions of dollars in damage and are one of the leading causes of weather-related deaths in the United States each year. Tornadoes are known to frequently form in the warm sector of extratropical cyclones (ETCs), yet relatively little research exists quantifying the climatological relationship between the two phenomena and their covariability. This work analyzes the climatology of F/EF1+ tornadoes relative to ETCs (ETCTORs) using historical databases of tornadoes and hourly ETC centers for 1980-2022. Most tornadoes (72\%) occurred broadly to the southeast of an ETC center within 2000km, with a median distance of approximately 500km. Of those tornadoes, 69\% occurred in outbreaks of 6+ tornadoes. The spatial and ETC-relative distributions are similar across all intensity levels. Through the seasonal cycle, tornadoes shift north and south along with ETCs and the jet stream, and they are much more strongly (weakly) associated with ETCs in the winter (76\%; summer 38\%) when the jet stream and ETCs are strongest (weakest). Finally, median tornado and tornado-producing ETC locations covary strongly with one another interannually. A long-term eastward trend in tornado-producing ETCs mirrors the eastward trend in tornadoes found in recent work, suggesting that this shift may be driven by spatial shifts in ETC tracks. This work identifies a strong link between tornado and ETC activity over the U.S. that warrants deeper analysis of their role in generating tornadic environments in a warming world.}

\begin{document}

\maketitle

%
%
%
\statement
	 Tornadoes are known to often form in the warm sector of the low pressure systems called ``extratropical cyclones'' that produce much of the rainfall over the continental United States. The relationship between the two hasn't received much research attention though. This study finds a strong relationship between the two spatial, seasonal, and interannually. This relationship may explain why tornado activity has been found to be shifting southeastward over the past few decades, and it may offer a pathway to better predict how tornado activity, which is small scale and difficult to model directly, may change with global-scale climate change via changes in the jet stream and the extratropical cyclones that form and track along it. 

%
%

%



\section{Introduction}
Tornadoes are one of the leading causes of weather-related fatalities in the United States and cause approximately \$1 billion dollars in normalized damage each year on average \citep{simmons2013,agee2019}. While there is a long-term decline in tornado fatalities owing to advances in forecasting technology and early-warning systems \citep{agee2019}, population growth and urban expansion continue to rapidly increase tornado exposure \citep{ashley2014, ashley2016, strader2024}. In addition to societal changes, climate change poses new threats for regions at risk. For instance, models \citep{sobel2011,lepore2021} have shown increases in convective available potential energy (CAPE) -- a key environmental factor for severe thunderstorms and tornadoes -- within a warming climate. Recent trends indicate that tornadoes have increasingly clustered into larger outbreaks in the US \citep{tippett2016}. In addition, spatial-temporal trends have shown changes in tornado frequency across various portions of the US, including increased activity in the Midwest and Southeast and decreased activity broadly across the Great Plains \citep{gensini2018,graber2024a,coleman2024}, perhaps consistent with expected shifts in severe thunderstorms in general \citep{ashley2023}. It is important to understand how tornadoes will change in the future and to consider all factors that contribute to severe weather and tornado formation.



Research in the middle of the 20th century first identified common synoptic patterns associated with severe thunderstorms and tornadoes  \citep{miller1959, doswell1980}. This work identified key synoptic drivers, including mid-level troughs, surface lows associated with extratropical cyclones (ETCs; a.k.a mid-latitude cyclones), and low-level jets that enhance the advection of warm, moist air inland from the Gulf of Mexico and also generate strong tropospheric wind shear. Research beginning in the 1980s demonstrated that tornadoes typically form from a pre-existing severe thunderstorm, whose formation requires as ingredients relatively high values of CAPE and lower tropospheric (0-6km) wind shear \citep{weisman1982,brooks2003} in order to sustain a long-lived rotating updraft. A third ingredient of strong near-surface (0-1 km) wind shear is required for tornado formation. This wind shear enhances SRH, providing a source of rotation at the surface \citep{markowski1998a,grams2012,coffer2019,malloy2024}. Recent work has demonstrated that tornadoes often form in the warm sector of an ETC between the surface cold and warm fronts \citep{meteotoday1994}, which provides an especially favorable environment for tornadoes where necessary ingredients overlap \citep{mercer2012,tochimoto2016,tochimoto2019,tochimoto2022}. Near-surface air mass boundaries, such as these, were found to be associated with 70\% of tornadoes and function as convergence zones where moisture and temperature gradients intensify, initiating convection and enhancing vertical wind shear \citep{maddox1980, markowski1998b}. Given that tornadoes depend specifically on the thermodynamic structure near the surface, including warm, moist air and 0-1km lifted condensation level (LCL) heights which increase storm intensity \citep{craven2004}, the ETC surface cyclone is an especially relevant synoptic feature to study. 

Climatologically, tornadoes and tornadic environments are most commonly found downstream of the Rocky Mountains  \citep{brooks2003,taszarek2020,li2020} and poleward of the warm and smooth Gulf of Mexico which reduces friction from surface roughness and enhances the transfer of low-level moisture \citep{li2021,li2024}. This region, primarily over the eastern half of the United States, is the principal hotspot for tornado activity globally \citep{maas2024}. This same region is also a hotspot for ETC activity over North America \citep{bentley2019}, because ETC formation is locally enhanced over the northern Great Plains just east of the front range of the Rocky Mountains \citep{fritzen2021}. Recent studies have found evidence of an eastward shift in tornado activity in the US over the past 40 years \citep{gensini2018}, though the reason for this shift remains unclear, including whether it is due to interannual variability or long-term warming. Given the known linkages that ETCs produce favorable environments and tornadoes often form in their warm sector, changes in ETC activity could play a role in this shift, and they may mediate the known linkage between variations in synoptic jet-stream patterns and tornado activity \citep{elkhouly2023,tippett2024,graber2024b,jiang2025}. Moreover, recent studies have found a decreasing trend in intense (EF2+) tornado activity and a broadening of the tornado season both temporally and spatially \citep{zhang2023,graber2024a,coleman2024}, with tornadoes concentrating into larger outbreaks \citep{tippett2016}. ETCs may provide an intermediate scale bridge to better understand how tornadoes, which are too small to directly simulate in modern weather and climate models, may change with climate variability (e.g., ENSO \citep{allen2015}) and long-term climate change \citep{tippett2015,allen2018}, as well as in subseasonal forecasting \citep{gensini2020}.

These outcomes first require understanding of the climatological spatiotemporal relationship between tornadoes and ETCs, which does not currently exist. This study fills this gap by quantifying the spatial relationship between the tornado and ETC centers and its temporal variability across seasons and interannually. By combining historical tornado data, ETC center tracking data, and European Centre for Medium-Range Weather Forecasts (ECMWF) fifth generation reanalysis data (ERA5), this work seeks to answer the following research questions:
\begin{enumerate}
    \item What percent of tornadoes are associated with an ETC (ETCTORs)?
    \item Where do ETCTORs form relative to the ETC center?
    \item Do ETCTORs most often occur in larger outbreaks?
    \item Do more intense tornadoes occur closer to the ETC center?
    \item How does the relationship between ETCs and tornadoes vary seasonally? Annually?
\end{enumerate}
The motivation of this paper is to provide a foundation for deeper meteorological analysis of how ETCs generate these tornadic environments in the future. Additionally, we note that this study neatly complements recent studies examining the climatology of tornadoes produced by landfalling tropical cyclones \citep{schultz2009, edwards2012, paredes2021} as well as during extratropical transition \citep{crosby2021}, thereby spanning tornado activity across the two principal types of surface cyclones in our climate.

Section 2 describes our datasets and methodology. Section 3 presents our results for the full climatology as well as decomposed by outbreak size, tornado intensity, season, interannually, and finally long-term trends. Section 4 summarizes results and provides discussion of key conclusions and avenues of future work.

\section{Data and Methods}
\subsection{Datasets}
This study uses a combination of datasets to analyze the climatology of tornadoes relative to ETCs from 1980 to 2022. Data is less reliable prior to 1980 \citep{NWS2010}.
\subsubsection{Tornado data}
For historical tornadoes, we use the NOAA's Storm Prediction Center (SPC) publicly-available single-track tornado dataset file (referred to as OneTor) which includes one entry per individual tornado from 1950 to 2022 (1950-2022\_actual\_tornadoes.csv) found at  \url{https://www.spc.noaa.gov/wcm/#data} (version updated 25 April 2023). All tornado times in OneTor are in CST (except 6 unspecified cases that are not used), but have been converted to UTC to match other datasets used herein.

Tornado intensity is based on the F/EF scale, which uses the Fujita (F) scale prior to 2007 and the Enhanced Fujita (EF) scale thereafter. This change in report ranking resulted in large increases in weak (F/EF0) tornado reports \citep{verbout2006}, and therefore we exclude these tornadoes from our dataset as is commonly done in tornado research \citep{zhang2023, coleman2024}. Tornadoes of unknown intensity are also removed from our dataset \citep{edwards2021}. A few landspouts may also be included in the dataset although most are likely omitted with the removal of weak F/EF0 tornadoes \citep{doswell1993}. 

About 3.4\% of tornadoes \citep{schultz2009} are generated by tropical cyclones, occasionally moving inland principally over the southeastern US \citep{schenkel2020}. Although most are likely omitted based on their F/EF0 rating, we explicitly remove these tornadoes using two different databases of tropical cyclone tornadoes (TCTOR): 1) for 1995 onwards we use the TCTOR dataset of \citep{edwards2012, edwards2022} from NOAA's Storm Prediction Center (SPC; \url{https://www.spc.noaa.gov/exper/tctor/}; accessed: 20 May 2024); 2) for 1982-1994 we use the TCTOR dataset from \citet{schultz2009}. For each F/EF1+ TCTOR, we removed those tornadoes that occurred in the same hour and within 2.5 degrees longitude and latitude of the entry. The result removes 776 F/EF1+ tornadoes from the tornado dataset for 1980-2022. 
We also note that due to missing TCTOR data prior to June 1982, it is possible that a small number (likely $<$10) of EF1+ tornadoes in our dataset could still be associated with a tropical cyclone.


\subsubsection{ETC tracking data}
\citet{gore2023} created a climatological database of ETC center positions in the northern hemisphere from the ERA5 Reanalysis database (described below) based on the location of minimum sea-level pressure using the TempestExtremes tracking software \citep{ullrich2021} at 6-hourly timesteps. We repeated this analysis at hourly timesteps to yield an hourly ETC center database to match the temporal resolution of the standard ERA5 output.

Since this study is focused on tornadoes within the continental US, we filter the ETC tracking to fit in our domain, extending from 135W to 55W in longitude and from 20N to 70N in latitude.

\subsubsection{ERA5 reanalysis data}
Hourly near-surface and pressure level data from the European Center for Medium-Range Weather Forecasting ERA5 Reanalysis dataset \citep{Hersbach2020} at 0.25 degree grid spacing for the period 1980-2022 are accessed from the National Center for Atmospheric Research (NCAR) Research Data Archive \citep{era5} at \url{https://doi.org/10.5065/BH6N-5N20}.

\subsection{Methods}
At each UTC hour during the period 1980-2022, we first check if a tornado occurs within the hour (e.g. at hour 14 we check if a tornado occurred between 1400-1459). Although multiple tornadoes may occur in a single hour, each tornado is analyzed as an individual data point. For each tornado occurrence, we extract contemporaneous ETC track points. If multiple ETC tracks exist, we calculate the distances between the tornado and each ETC track point and choose the closest point, so the tornado is matched with only one ETC track. Some tornado cases have no ETC present within the ETC domain and are labeled as such.

\subsubsection{Defining ETC-associated tornadoes (ETCTORs)}
Given that one of our goals is to examine where tornadoes occur relative to an ETC center, we want to identify tornado cases that are likely associated with an ETC (ETCTOR). We do so via a simple proximity filter meant to capture the basic geometry of the ETC warm sector: tornadoes that occur within a distance of 2000 km in the southwest, southeast, and northeast quadrants relative to the ETC low pressure center, and within a shorter distance of 500 km in the northwest quadrant. All other tornadoes are deemed not associated with an ETC (``non-ETC''), which includes tornadoes that are both too far from an ETC center (i.e. do not meet the above proximity criteria) and that have no ETC within the domain. Below we detail our motivation for this approach. 

\begin{figure}[t]
    \centering
    \noindent\includegraphics[width=1\linewidth]{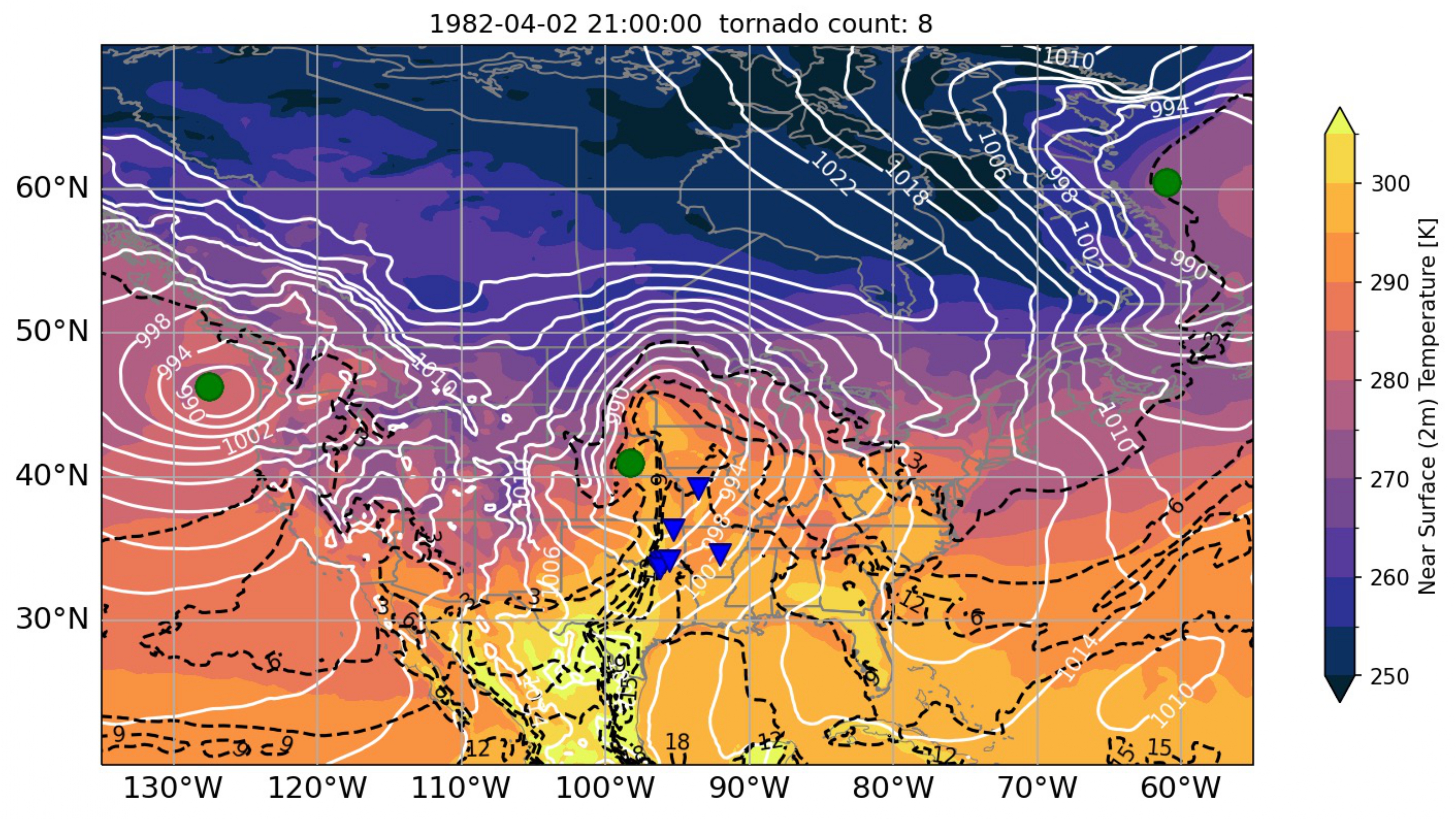}
    \caption{Example plot of a tornado hour with mean sea-level pressure [hPa] (white contours), specific humidity [g/kg] (black dashed contours), temperature [K], (colors), tornado point(s) (blue triangles), and ETC track point(s) (green dots).}
    \label{fig:domain}
\end{figure}

We used the ERA5 near-surface 2-meter temperature, specific humidity, and the mean sea-level pressure to identify the ETC warm sector and near-surface circulation of the ETC center. To illustrate our methodology, we begin with an example case from 21 UTC on 2 April 1982 when multiple tornadoes (8) and ETCs (3) exist within our domain (Figure \ref{fig:domain}). The ETC track centers from our ETC database are found at local minima of sea-level pressure and the tornado points are found to the southeast of the closest ETC track point, which is in eastern Nebraska. The tornadoes are clearly located in the warm sector of the cyclone, where there is abundant warm and moist air near the surface. Meanwhile, there are two other ETC centers in our domain that are not associated with these tornadoes and are ignored since we only match the tornadoes to the closest ETC.


\begin{figure}[t]
    \centering
    \includegraphics[trim={4cm 0 4cm 0},width=0.7\linewidth]{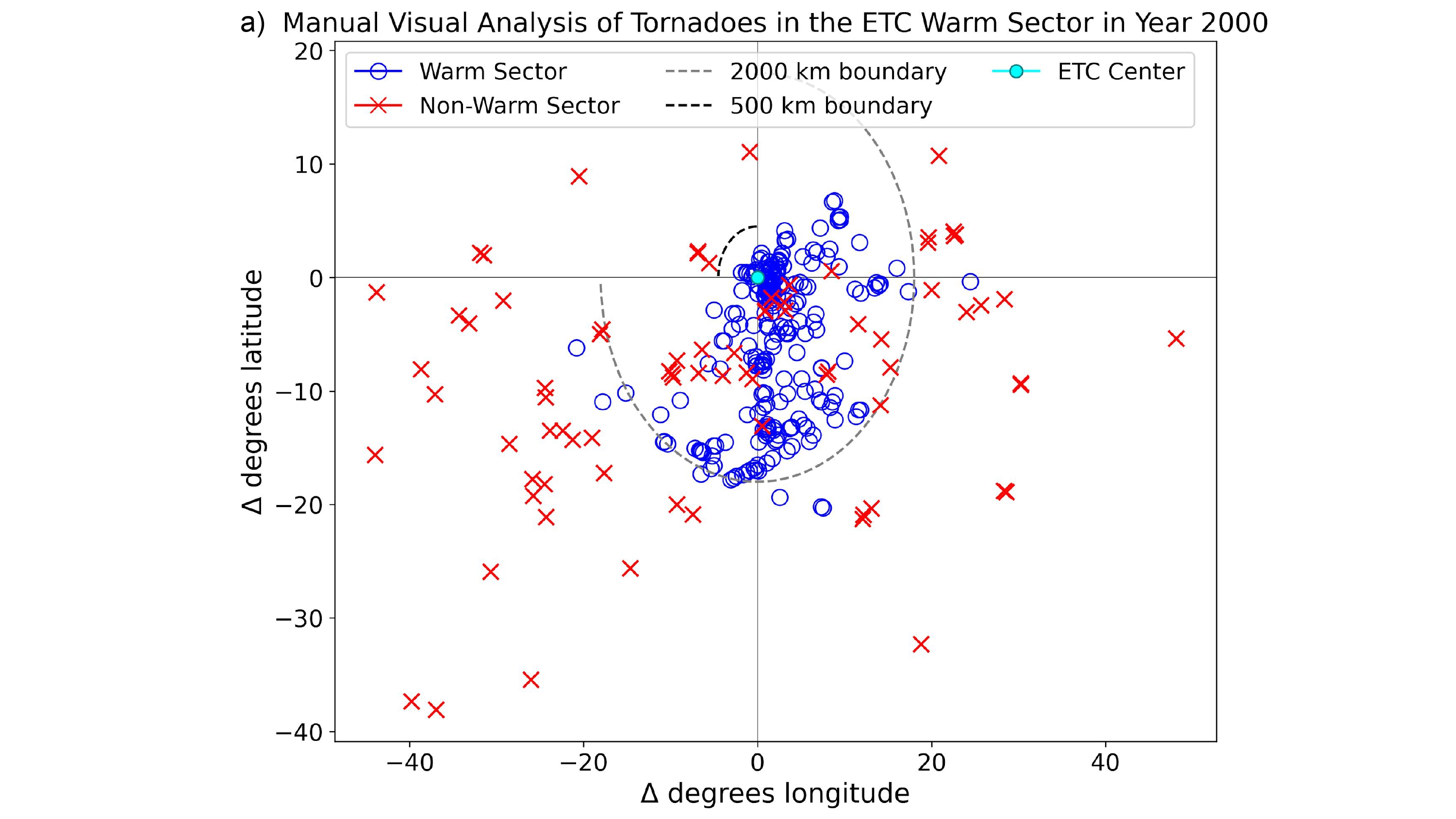}
    \noindent\includegraphics[width=1\linewidth]{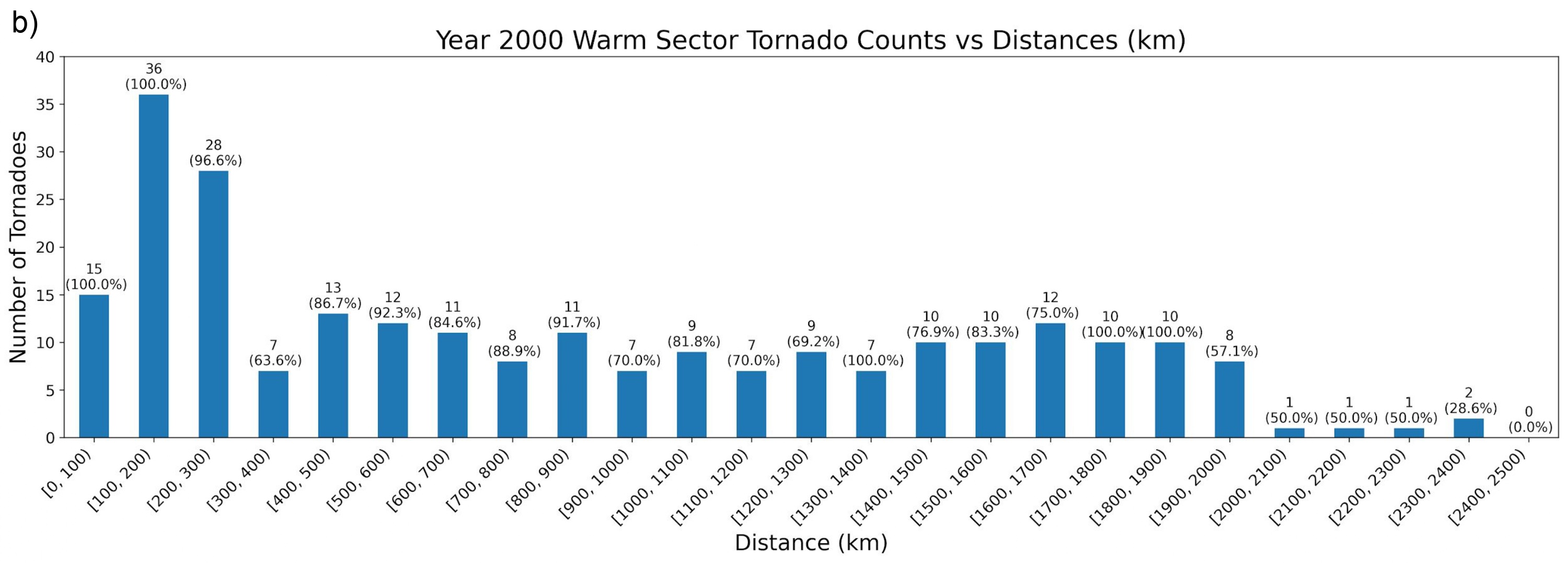}
    \caption{(a) Distribution of tornadoes in the ETC warm sector (blue circles) and non-ETC warm sector (red Xs) tornadoes relative to ETCs centered at (0,0) (cyan dot) in the year 2000, including the 2000 km distance threshold (grey dotted line) and the 500 km NW boundary. (b) Bar plot of the number and percentage of warm sector tornadoes (y-axis) found within 100 km bins of distances between a tornado and ETC (x-axis) in the year 2000.}
    \label{fig:assoc-2000}
\end{figure}

To define a simple and reasonable proximity filter for ETCTORs, we perform this visual analysis for all cases in the year 2000 and identify which cases are and are not in the warm sector of its paired ETC as a function of geographic location relative to ETC center in Figure \ref{fig:assoc-2000}a. (An excel file with table of data from this analysis is included as Supplementary.) Additionally, Figure \ref{fig:assoc-2000}b shows the number of tornadoes associated with an ETC, and the percentage of all tornadoes identified as associated, as a function of distance from ETC center. Tornadoes within 2000 km of an ETC were found to be nearly always within the warm sector of the ETC. Beyond 2000 km, there is a sharp decrease in both the total number of tornadoes and the percentage of those tornadoes in the warm sector. Hence, we choose 2000 km as a simple maximum distance threshold for a tornado to be deemed likely associated with an ETC. This distance has some physical intuition as it corresponds to the characteristic radius of the ETC circulation \citep{rhines1975} and hence its range of influence for producing a tornado. Our visual analysis indicates that the warm sector of an ETC is typically focused in the southeast quadrant of the ETC center but can extend into the northeast and southwest quadrants too. However, it does not extend into the northwest quadrant, though occasionally the ETC low pressure center may be shifted slightly southeast of the edge of the warm sector. For these reasons, we use the 2000 km threshold for the southeast, southwest, and northeast quadrants and use a much shorter 500 km threshold in the northwest quadrant. The choice of 500 km radius in the northwest quadrant is further based on the seasonal analysis below where we found that summer tornadoes more frequently extend up to 400 km to the northwest of the ETC center, but beyond 500 km they are quite rare. The resulting boundaries from the manual analysis of the year 2000 tornado cases are marked in Figure \ref{fig:assoc-2000}a, and this filtering has been applied to all tornadoes in our time period 1980-2022 to differentiate between ETCTORs and non-ETC tornadoes.

We note an ideal approach would be to automate this process, but our visual analysis made clear that the geometry of the warm sector is quite complicated from case to case and through time given that the cyclone evolution is superimposed onto a background state with strong pre-existing and time-varying temperature gradients, which makes automation difficult. This topic could be a great avenue for future work.

\subsubsection{Analysis methodology}
To analyze the geographic distributions of ETCTOR, we create spatial maps using 2D histograms with 0.5-degree bin spacing. The median latitude and longitude position of ETCTOR is used to define the central location of tornado activity, and the 25th and 75th percentiles are calculated to highlight the spread in tornado activity. In further analyses, similar spatial maps are created to analyze the spatial distributions of tornadoes as well as ETCs (depicted by contour lines).

\begin{figure}[ht]
    \centering
    \includegraphics[width=0.75\linewidth]{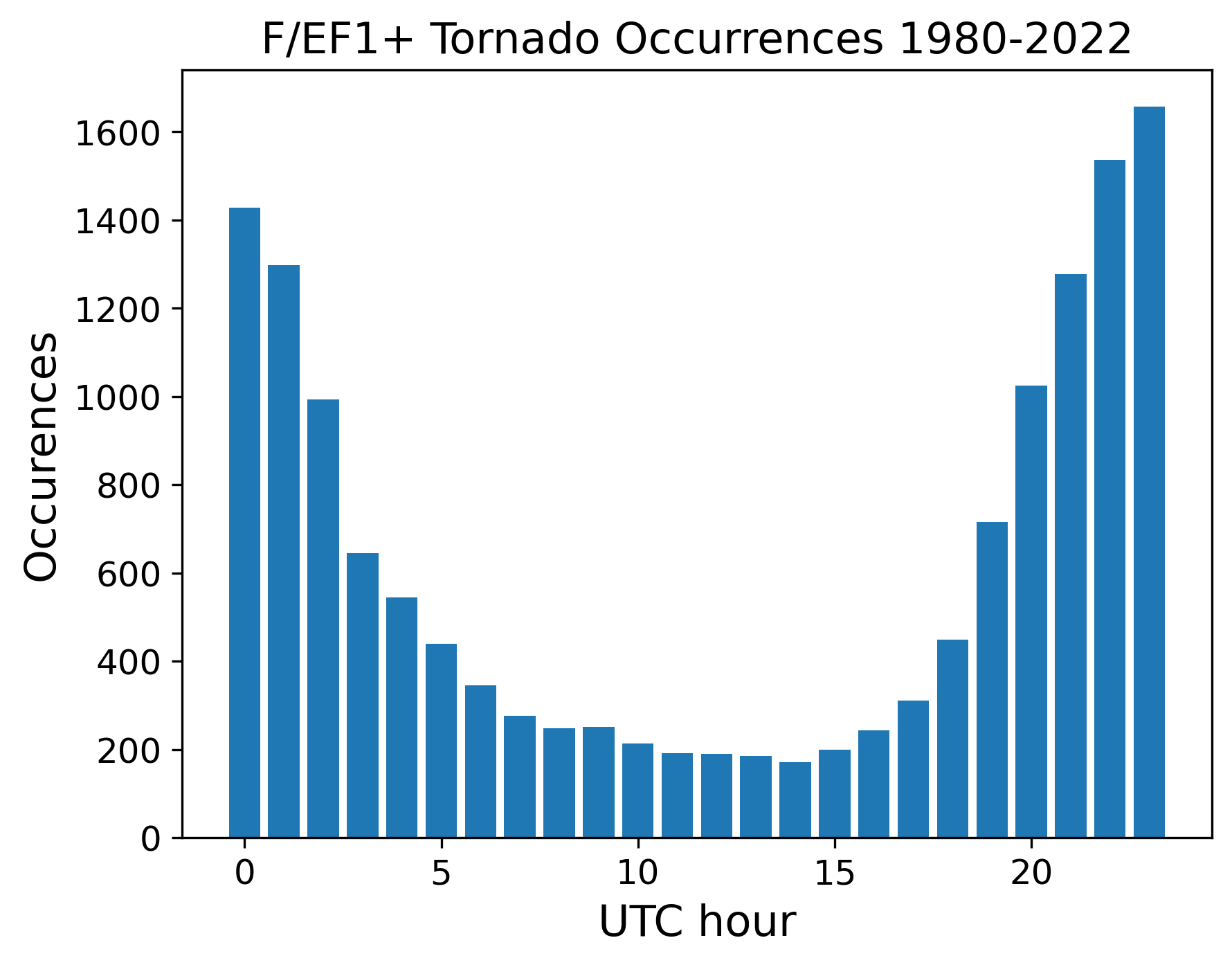}
    \caption{Climatological diurnal cycle of UTC hour of F/EF1+ tornado occurrences in the US during the time period 1980-2022.}
    \label{fig:cst-hour}
\end{figure}

For ETCTORs, we create analogous ETC-relative spatial maps using 2D histrograms of 1-degree bin spacing to analyze where tornadoes occurred relative to their closest ETC track. These datasets are also used to calculate the distance and direction of the tornado from the ETC center, which are visualized using wind rose plots for a clearer indication of the precise distances and cardinal directions. We then repeat these analyses across different dimensions of the ETCTOR dataset including outbreak size, intensity, season, and interannual variability and trends (Section \ref{sec:results}\ref{subsec:outbreak}-\ref{subsec:interannual}). The F/EF1+ tornado outbreak sizes are defined by the number of tornadoes in a tornado day: large (6 or more tornadoes) \citep{fuhrmann2014, tippett2016}, small (2-6 tornadoes) and isolated (one tornado). We define a tornado day as beginning and ending at 2:00pm UTC as this is the time of minimum tornado activity during the climatological diurnal cycle (Figure \ref{fig:cst-hour}); see also \citep{amanda2010}. 



\section{Results}\label{sec:results}

\subsection{Distribution of tornadoes associated with an ETC (ETCTORs)}\label{subsec:tornadoesassoc}

\begin{figure}
    \centering
    \text{Distribution of ETCTORs}\par\medskip
    \includegraphics[trim={4cm 0 0 0},width=1\linewidth]{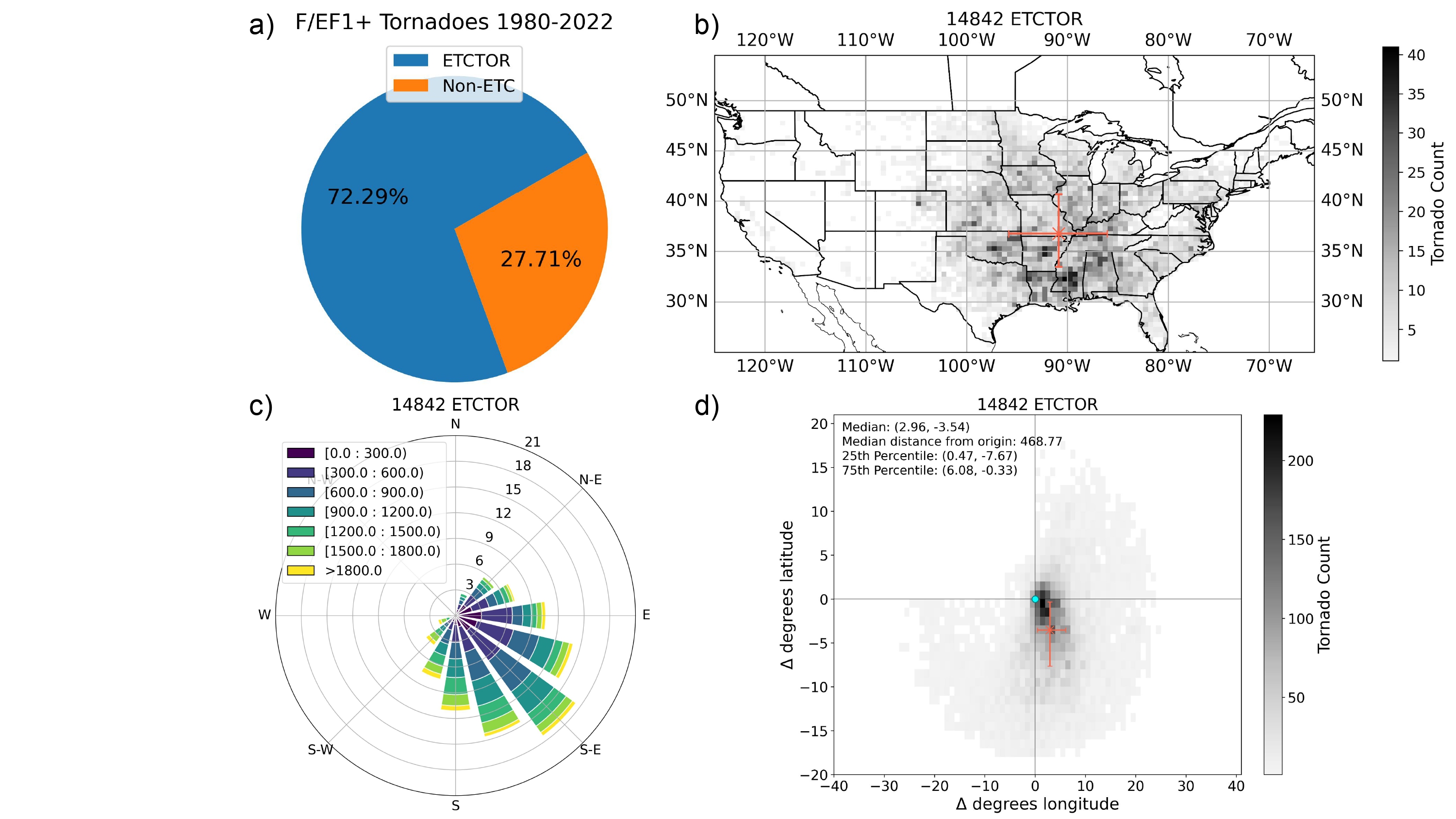}
    \caption{Frequencies and distributions of F/EF1+ tornado occurrences between 1980-2022. (a) percentage of ETCTORs and non-ETC tornadoes; (b) geographic distribution of ETCTOR (grey shading) including the median tornado position (red X) and the 25th and 75th percentiles (red bars); (c) distances and directions of ETCTOR relative to ETCs; (d) distribution of ETCTOR (grey shading) relative to ETCs (cyan dot) centered at (0, 0) including the median longitude and latitude of tornadoes (red X) and the 25th and 75th percentiles (red bars).}
    \label{fig:tor-assoc}
\end{figure}

We first examine the climatological frequency and spatial distribution of tornadoes that are associated with an ETC (Figure \ref{fig:tor-assoc}). We find approximately 72\% of all tornadoes are ETCTOR (Figure \ref{fig:tor-assoc}a). The southern Great Plains and the lower Mississippi river basin are the major hotspots for ETCTOR (Figure \ref{fig:tor-assoc}b). The median position for ETCTOR is located near the border of Arkansas and Missouri. 
The distribution of ETCTOR relative to ETCs is also displayed. The median position of ETCTOR is approximately 469 km to the southeast of the ETC center (Figure \ref{fig:tor-assoc}d), with the vast majority of cases located in the southeast quadrant (Figure \ref{fig:tor-assoc}c). Analogous plots for non-ETC tornadoes can be found in supplementary material.  Given our simple distance thresholds for defining an ETCTOR, it is possible a small number of tornadoes that are associated with the ETC warm sector but are very far from the center may have been excluded (see Supplementary Figure S1), which would slightly increase our estimate of 72\% ETCTOR.

\subsection{Spatial relationship between ETCTOR and their ETCs}
\subsubsection{Variations by outbreak size}\label{subsec:outbreak}
\begin{figure}
    \centering
    \text{Spatial Distribution of Tornadoes and ETCs}\par\medskip
    \includegraphics[trim={0 2cm 0 0}, width=1\linewidth]{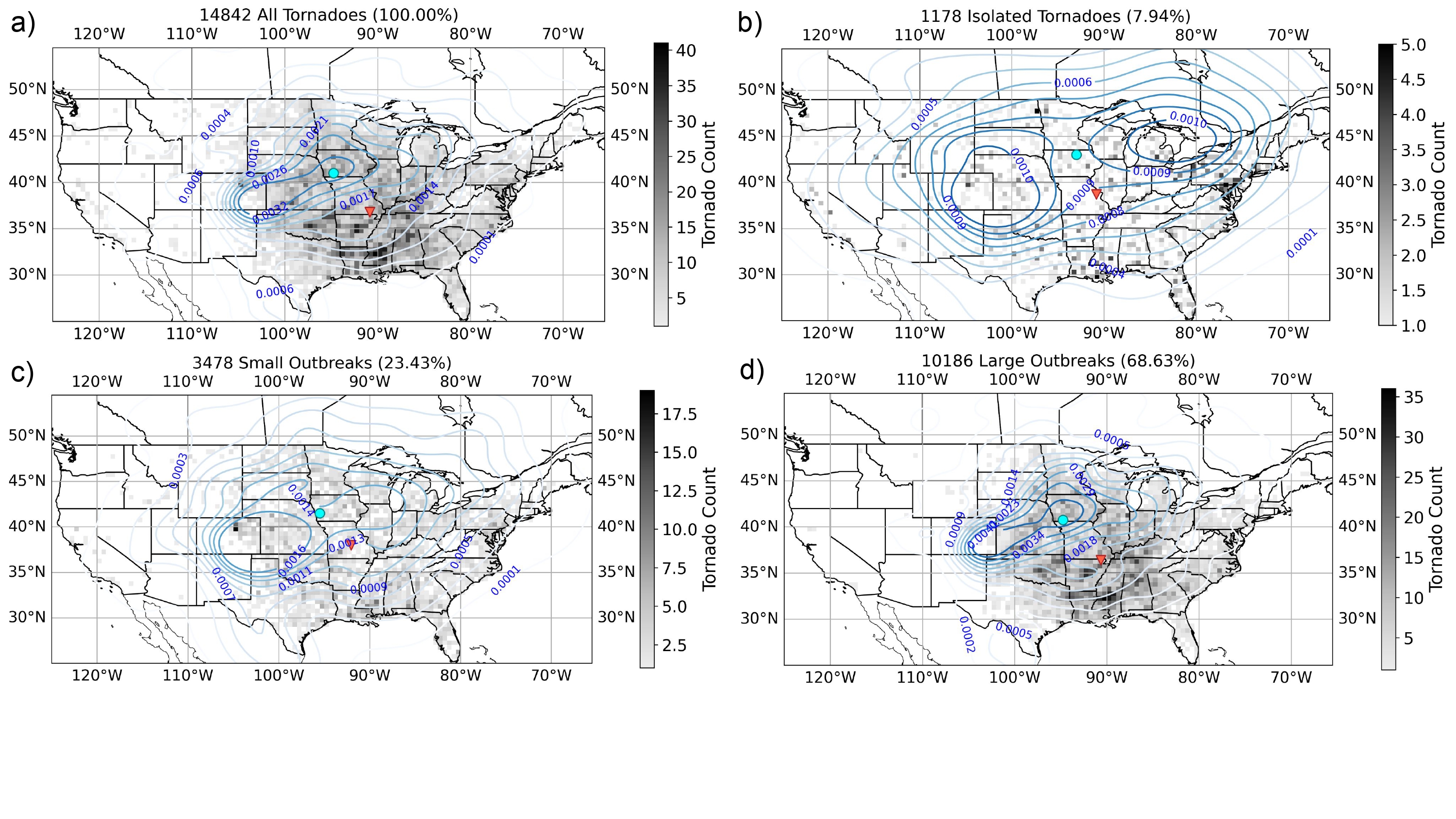}
    \caption{Geographic distribution of ETCTOR (shading) and their associated ETCs (contours) from 1980-2022 with the median tornado (red triangle) and median ETC (cyan dot) positions for (a) all tornadoes; (b) isolated tornadoes; (c) small outbreaks of 2-5 tornadoes; (d) large outbreaks of 6+ tornadoes. Contours show kernel density estimates (KDE) of ETC locations, with values representing spatial probability density in units of [deg$^{-2}$].}
    \label{fig:outbreak-geo}
\end{figure}

\begin{figure}
    \centering
    \text{Tornado Distribution Relative to ETC}\par\medskip
    \includegraphics[trim={7cm 0 0 0}, width=1\linewidth]{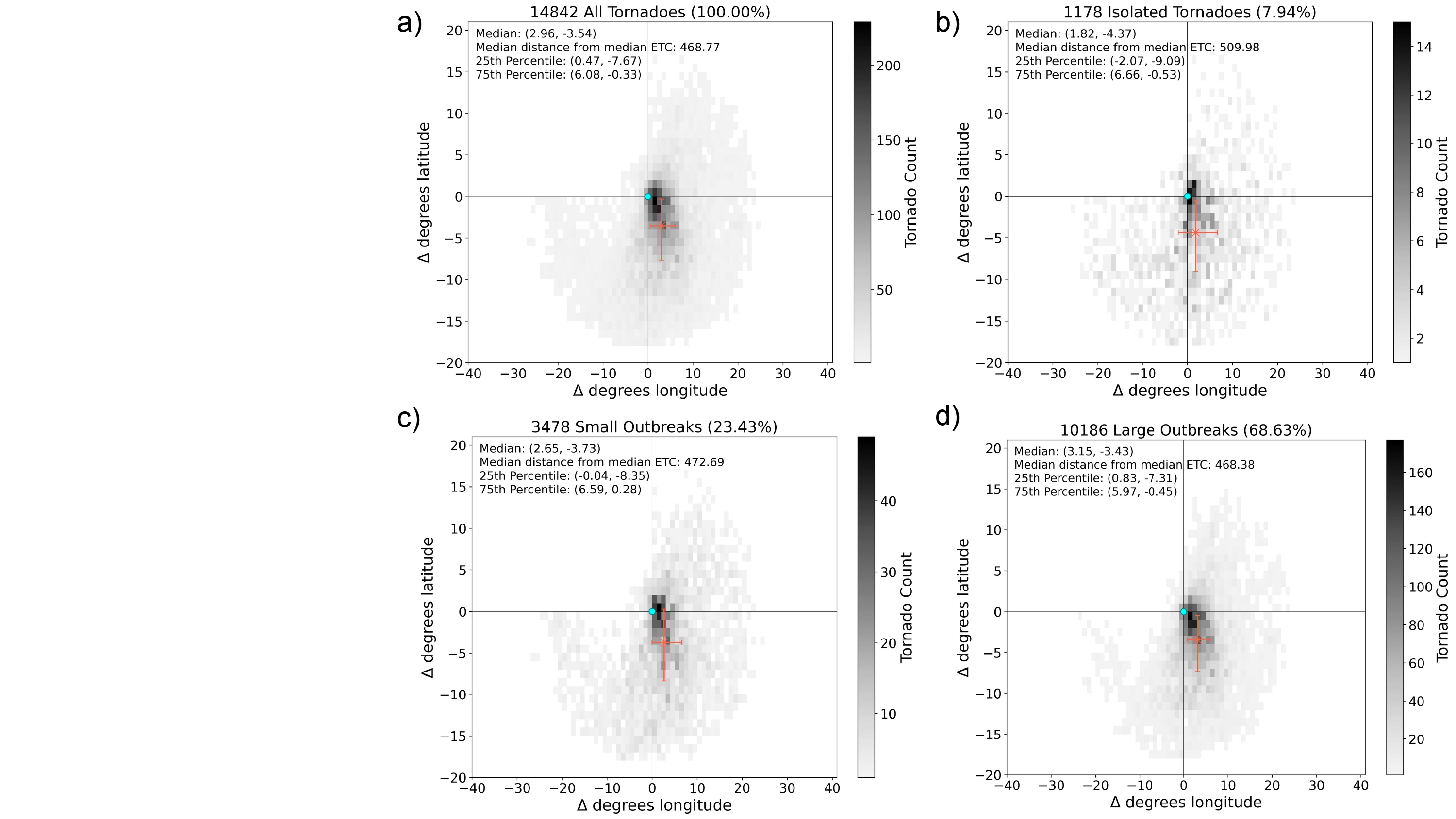}
    \caption{Distribution of ETCTOR (grey shading) relative to ETCs (cyan dot) centered at (0, 0) from 1980-2022 including the median longitude and latitude of tornadoes (red X) and the 25th and 75th percentiles (red bars) for (a) all tornadoes; (b) isolated tornadoes; (c) small outbreak of 2-5 tornadoes; (d) large outbreaks of 6+ tornadoes.}
    \label{fig:outbreak-dist}
\end{figure}

We next examine the spatial relationship between ETCTOR and ETCs and how this varies with outbreak size as defined in our analysis methodology: all tornadoes, isolated tornadoes, small outbreaks of 2-5 tornadoes, and large outbreaks of six or more tornadoes. The geographic distribution of all ETCTOR is shown in Figure \ref{fig:outbreak-geo}a. The ETC tracks are most common to the northwest of the primary tornado hotspot, with the corridor of peak frequency values extending from southeast Colorado through central Iowa and the median location in southwestern Iowa. The spatial distribution of tornadoes relative to the ETC center is shown in Figure \ref{fig:outbreak-dist}a (identical to Figure \ref{fig:tor-assoc}d), with median tornado location approximately 469 km southeast of the ETC center (approximately 3 degrees east and 3.5 degrees south) as discussed in the previous subsection, with an interquartile range of 0-6 degrees east of center and 0-8 degrees south of center. Tornadoes most frequently occur to the east and east-northeast very close to the ETC center within 200 km. Overall, ETCTORs tend to occur near the ETC center to the east/southeast but can often form farther away especially to the south of center.

Isolated tornadoes (approximately 8\% of all ETCTOR) have similar broad spatial distributions and median locations of tornadoes and ETCs to the all-tornado case. There are notable local tornado hotspots along the Gulf coast, especially in Louisiana and Florida, as well as in the northeast and Midwest (Figure \ref{fig:outbreak-geo}b). There are two broad local maxima of associated ETC frequency, one over western Kansas and eastern Colorado, and the other between Lake Michigan and Lake Huron. The median distance is approximately 10\% larger (507 km) to the southeast (approximately 2 degrees east and 4 degrees south) of the ETC center (Figure \ref{fig:outbreak-dist}b), and their relative location is more spread out (larger 25th and 75th percentile ranges).

Results for small outbreak tornadoes (approximately 23\% of all ETCTOR) are generally similar to isolated tornadoes except now shifted to the west towards the central Great Plains (Figure \ref{fig:outbreak-geo}c), with now just a single ETC frequency maximum centered over western Kansas and eastern Colorado. The median distance is approximately 483 km and shifted slightly eastward to be similar to the all-tornado case (Figure \ref{fig:outbreak-dist}c). The majority of tornadoes occur in large outbreaks (approximately 69\% of all ETCTOR), and therefore the spatial distributions of tornadoes in large outbreaks and ETCs are quite similar to the all-tornado case (Figure \ref{fig:outbreak-geo}d), with a median distance of approximately 468 km (Figure \ref{fig:outbreak-dist}d). 


\begin{figure}[t]
    \centering
    \text{Directional Distribution of Tornadoes Relative to ETC}\par\medskip
    \includegraphics[trim={5cm 0 5cm 0}, width=1\linewidth]{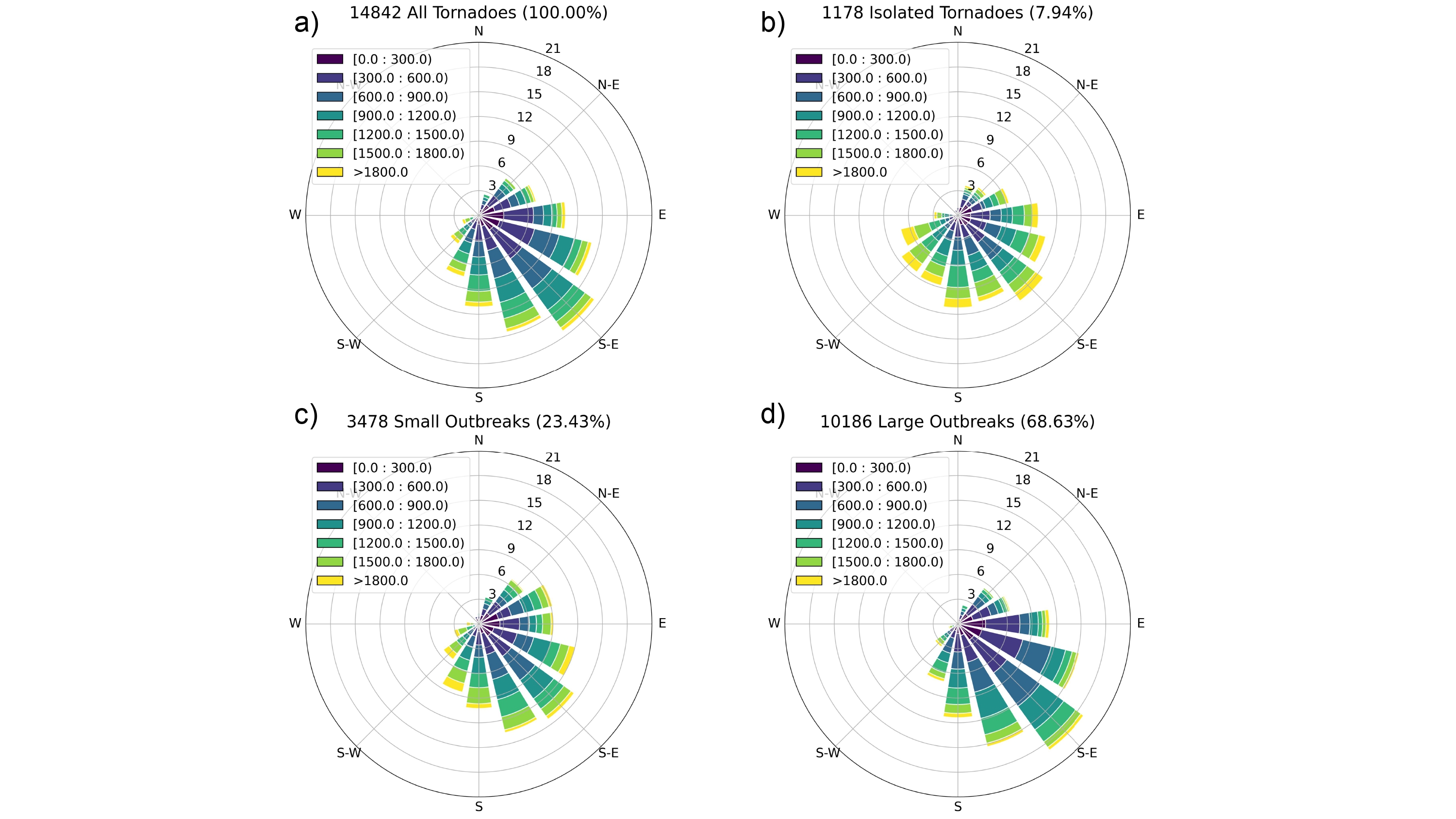}
    \caption{Distance and direction of ETCTOR relative to ETC track by outbreak size: (a) all tornadoes; (b) isolated tornadoes; (c) small outbreaks of 2-5 tornadoes; and (d) large outbreaks of 6+ tornadoes. Colors indicate distances from center and ring values indicate the frequency of occurrence.}
    \label{fig:wind-outbreak}
\end{figure}

Across analyses, larger outbreaks tend to cluster more tightly together in space relative to the ETC center as shown in the wind rose plots of Figure \ref{fig:wind-outbreak}d, which highlight the detailed directional distribution of tornadoes relative to ETC center that is difficult to assess in the previous figure.  Isolated tornadoes are more uniformly distributed towards the south (Figure \ref{fig:wind-outbreak}b) and over the full range of distances from 0-2000 km. As outbreak size increases (Figure \ref{fig:wind-outbreak}b,c,d), the directional range systematically clusters towards the southeast and distances shift closer to the ETC center. For example, approximately 19\% of large outbreak tornadoes occur due southeast of ETC center as compared to only approximately 12\% and 14\% for isolated and small outbreaks, respectively.

\subsubsection{Variations by tornado intensity}\label{subsec:EF}
\begin{figure}
    \centering
    \text{Spatial Distribution by Tornado Intensity}\par\medskip
    \includegraphics[trim={0 0 0 0}, width=0.75\linewidth]{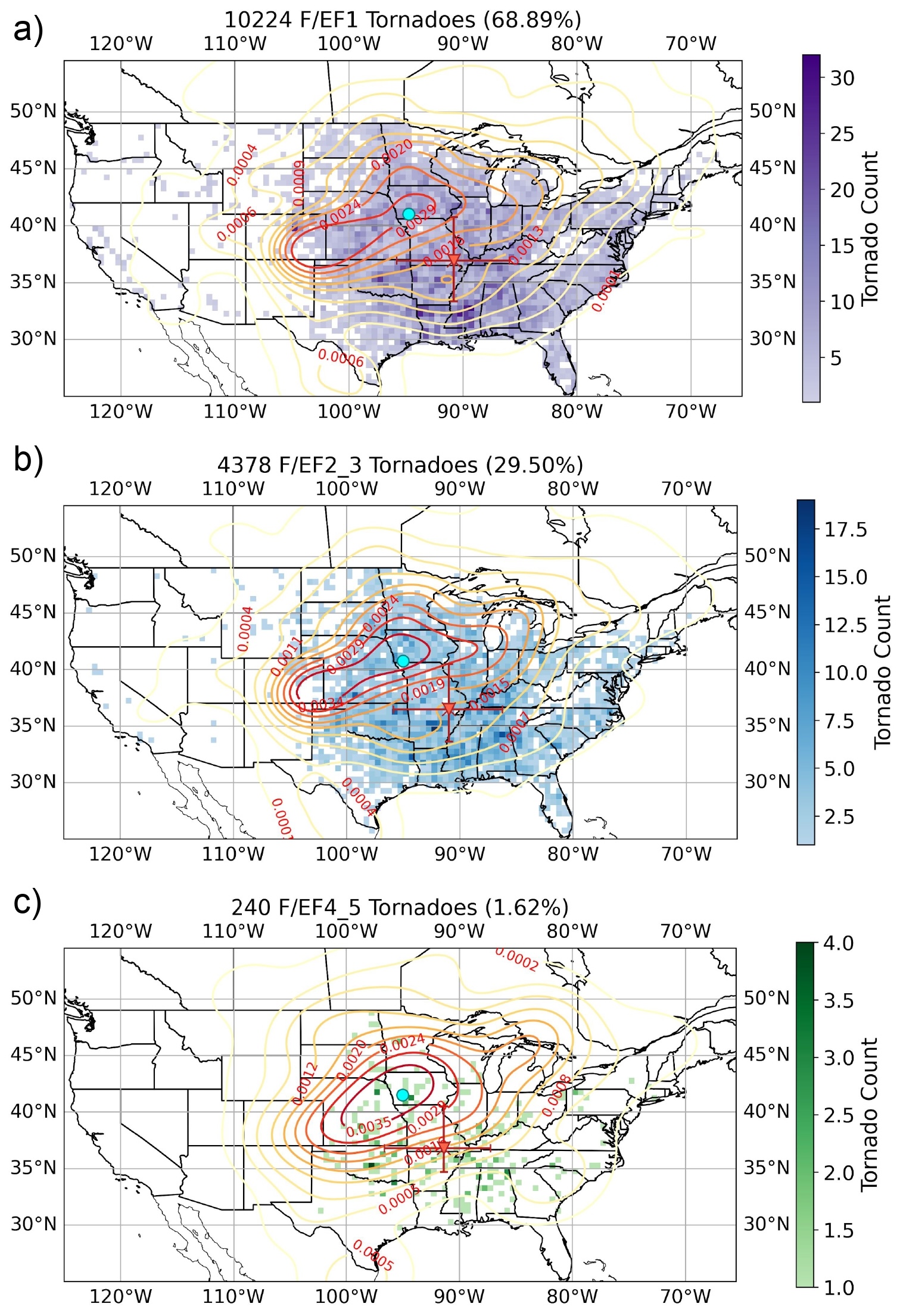}
    \caption{Geographic distribution of ETCTOR (shading) and their associated ETCs (contours) with the median tornado position (red triangle), the tornado 25th and 75th percentiles (red bars), and the median ETC position (cyan dot) by F/EF level: (a) F/EF1 tornadoes; (b) F/EF2-3 tornadoes; and (c) F/EF4-5 tornadoes.}
    \label{fig:geo-ef}
\end{figure}

\begin{figure}[t]
    \centering
    \text{Relative Distribution by Tornado Intensity}\par\medskip
    \includegraphics[trim={7cm 0 0 0}, width=1\linewidth]{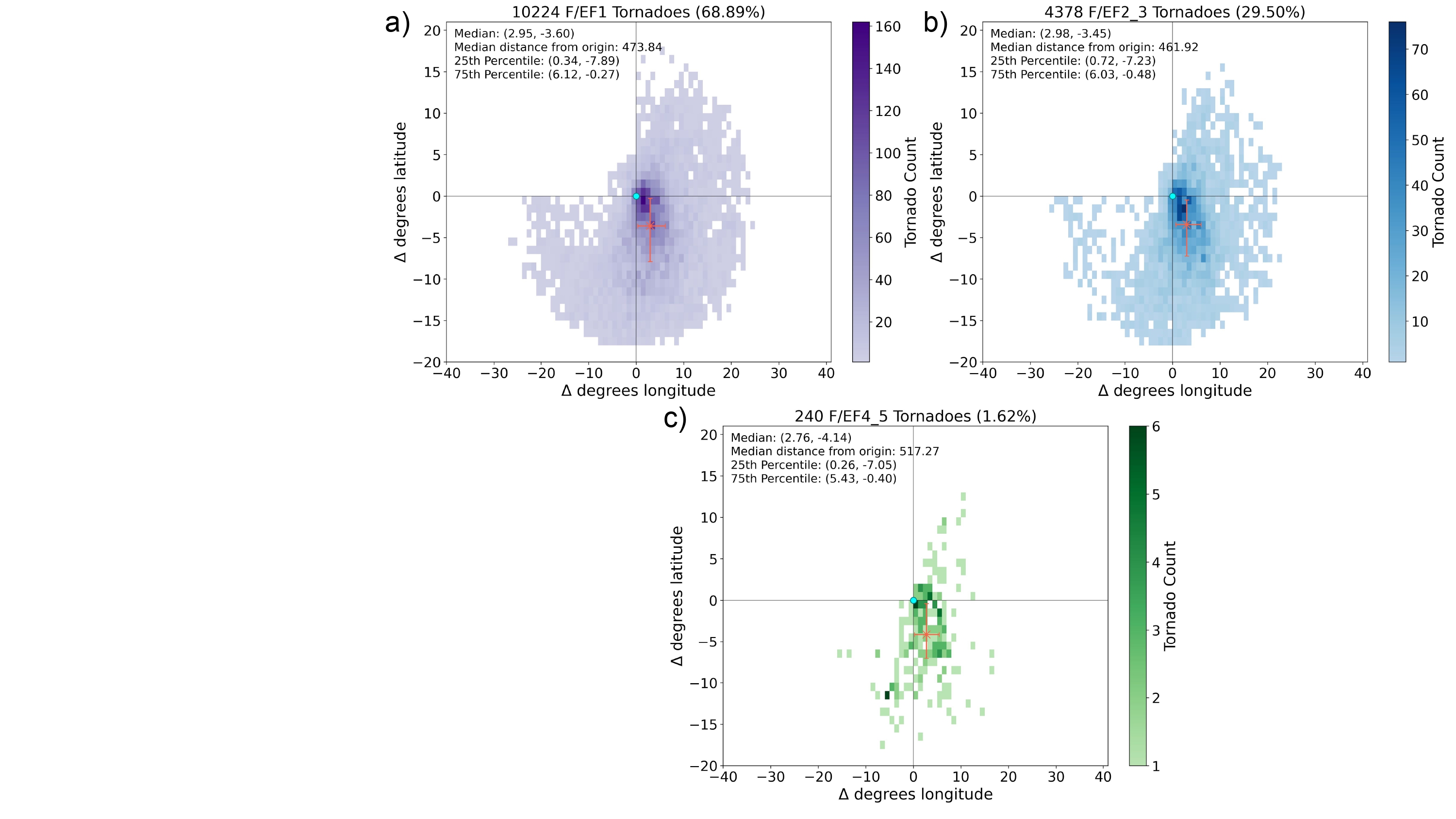}
    \caption{Distribution of ETCTOR relative to ETC center (blue dot) including the median longitude and latitude (red X) and the 25th and 75th percentiles (red bars) by F/EF level: (a) weak (F/EF1) tornadoes; (b) strong (F/EF2-3) tornadoes; and (c) violent (F/EF4-5) tornadoes.}
    \label{fig:distr-ef}
\end{figure}

The spatial distribution of ETCTOR and ETCs (Figure \ref{fig:geo-ef}) are quite similar across all tornado intensities and similar to all tornadoes (Figure \ref{fig:outbreak-geo}a). Although the median positions of ETCTOR and ETCs are similar across intensities, strong and violent tornadoes (Figure \ref{fig:geo-ef}b,c) tend to be located slightly farther inland based on the shortening of the southward interquartile range bar with increasing intensity. This may indicate a potential difference in distribution though it may also be due to the much smaller sample size for violent tornadoes. Weak tornadoes make up the majority of all tornadoes (approximately 69\%), while violent tornadoes make up less than 2\%. Tornado location relative to ETC center is also nearly constant across intensity (Figure \ref{fig:distr-ef}). The above outcomes may simply reflect the fact that tornado outbreaks that produce strong tornadoes also produce many weaker tornadoes too \citep{elsner2014}, so variations in the relationship occur at the outbreak scale and will not be evident for individual tornadoes of varying intensity.

\subsubsection{Variations by season}\label{subsec:season}


\begin{figure}
    \centering
    \text{Spatial Distribution by Season}\par\medskip
    \includegraphics[trim={0 2cm 0 0}, width=1\linewidth]{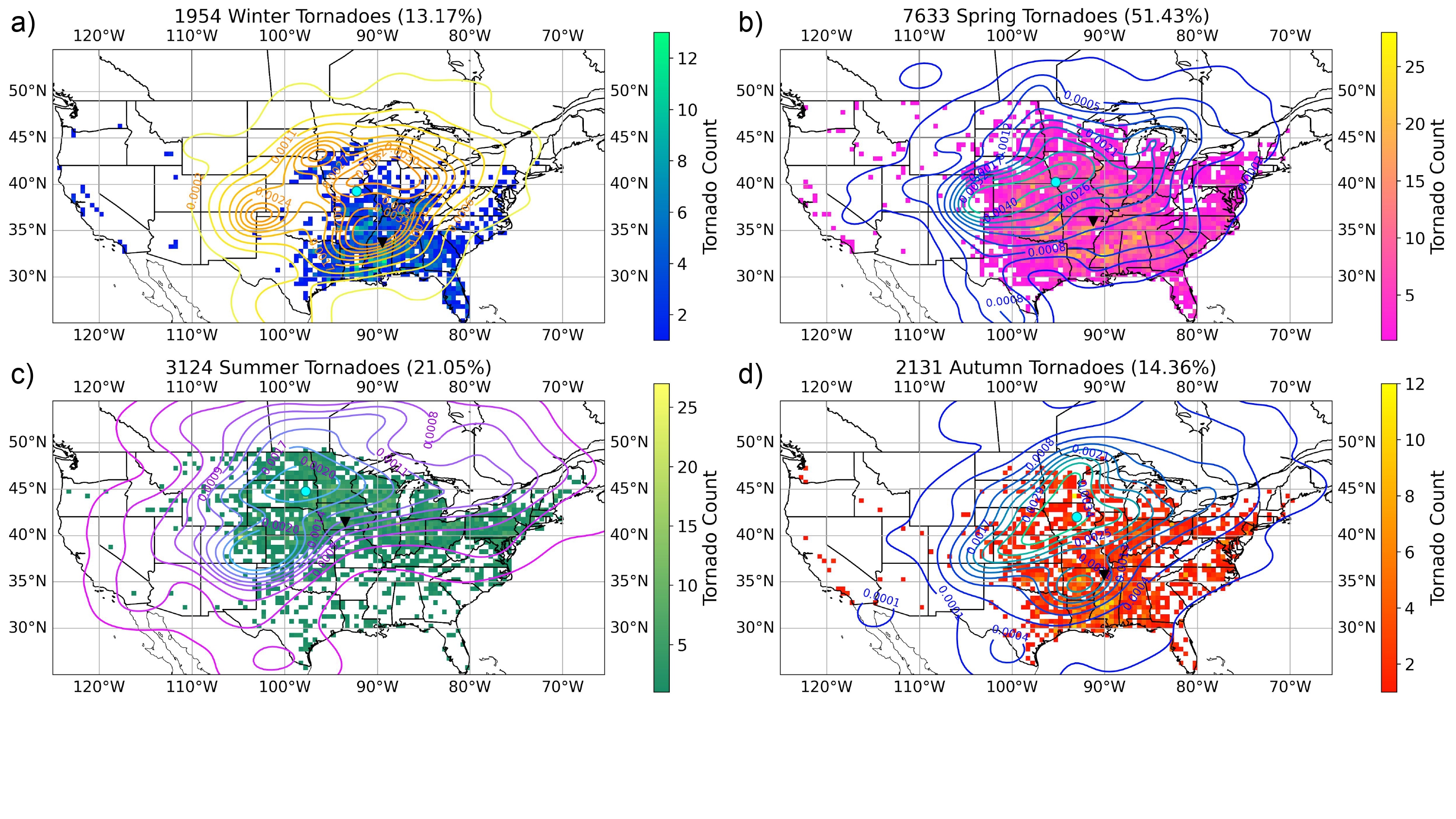}
    \caption{Geographic distribution of ETCTOR (color shading) and their associated ETCs (color contours) from 1980-2022 with the median tornado (black triangle) and median ETC (cyan dot) positions for (a) winter tornadoes; (b) spring tornadoes; (c) summer tornadoes; (d) autumn tornadoes.}
    \label{fig:season-geo}
\end{figure}

\begin{figure}
    \centering
    \text{Relative Distribution by Season}\par\medskip
    \includegraphics[trim={7cm 0 0 0}, width=1\linewidth]{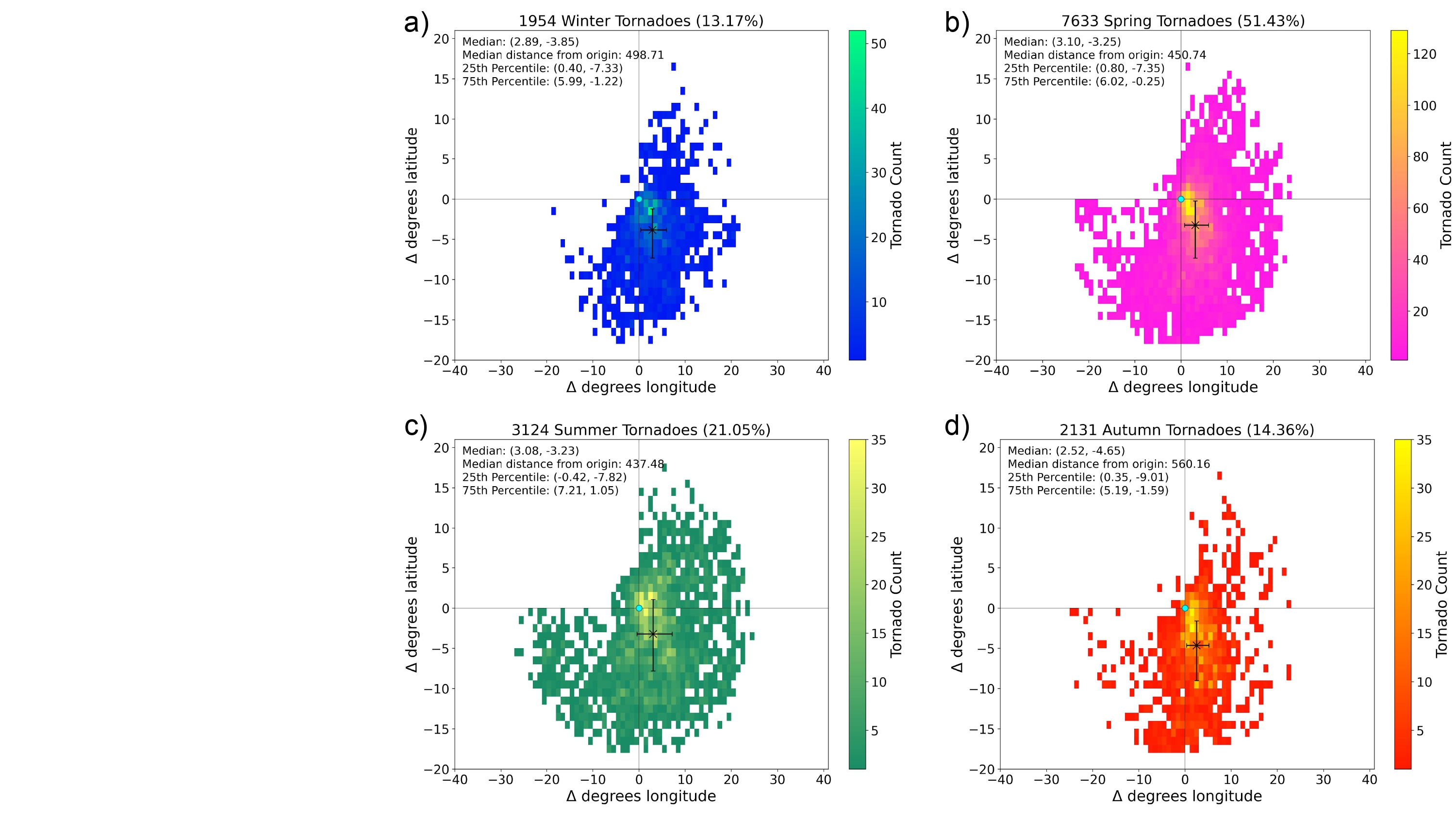}
    \caption{Distribution of ETCTOR (color shading) relative to ETCs (cyan dot) centered at (0, 0) from 1980-2022 including the median longitude and latitude of tornadoes (black X) and the 25th and 75th percentiles (black bars) for (a) winter tornadoes; (b) spring tornadoes; (c) summer tornadoes; (d) autumn tornadoes.}
    \label{fig:season-dist}
\end{figure}

The seasonal cycle of the spatial distribution of ETCTOR and ETCs (Figure \ref{fig:season-geo}) shifts northward and westward toward the continental interior during spring and summer and then back southward and eastward towards the Gulf Coast during autumn and winter. In winter, the median ETC location is in north-central Missouri and the median tornado location is in north-central Mississippi (Figure \ref{fig:season-geo}a). In summer, these shift to eastern South Dakota and south-central Iowa, respectively (Figure \ref{fig:season-geo}c). These shifts correspond closely to the seasonal cycle of the jet stream itself moving poleward as the hemisphere warms and equatorward as the hemisphere cools \citep{stendel2021}. 

In winter (13.17\% of all ETCTOR; Figure \ref{fig:season-geo}a), when the jet stream is farthest south and tornadoes cluster in the southeast, the ETC density distribution is shifted to the east with a hotspot over the Midwest from Tennessee and Arkansas northward through Illinois. In spring (51.43\% of all ETCTOR; Figure \ref{fig:season-geo}b), the distribution closely mirrors that for all tornadoes (Figure \ref{fig:outbreak-geo} a), since spring is the dominant season for tornadoes. In summer (21.05\% of all ETCTOR; Figure \ref{fig:season-geo}c), tornadoes shift strongly into the northern half of the United States north of 40N and the local maximum of associated ETCs is located over the Dakotas. Finally, in autumn (14.36\% of all ETCTOR; Figure \ref{fig:season-geo}d) the tornado distribution is comparably broad as in spring with a hotspot in the southeast and lower overall frequency.

\begin{figure}
    \centering
    \text{Relative Distribution by Season}\par\medskip
    \includegraphics[trim={0 0 0 0}, width=0.7\linewidth]{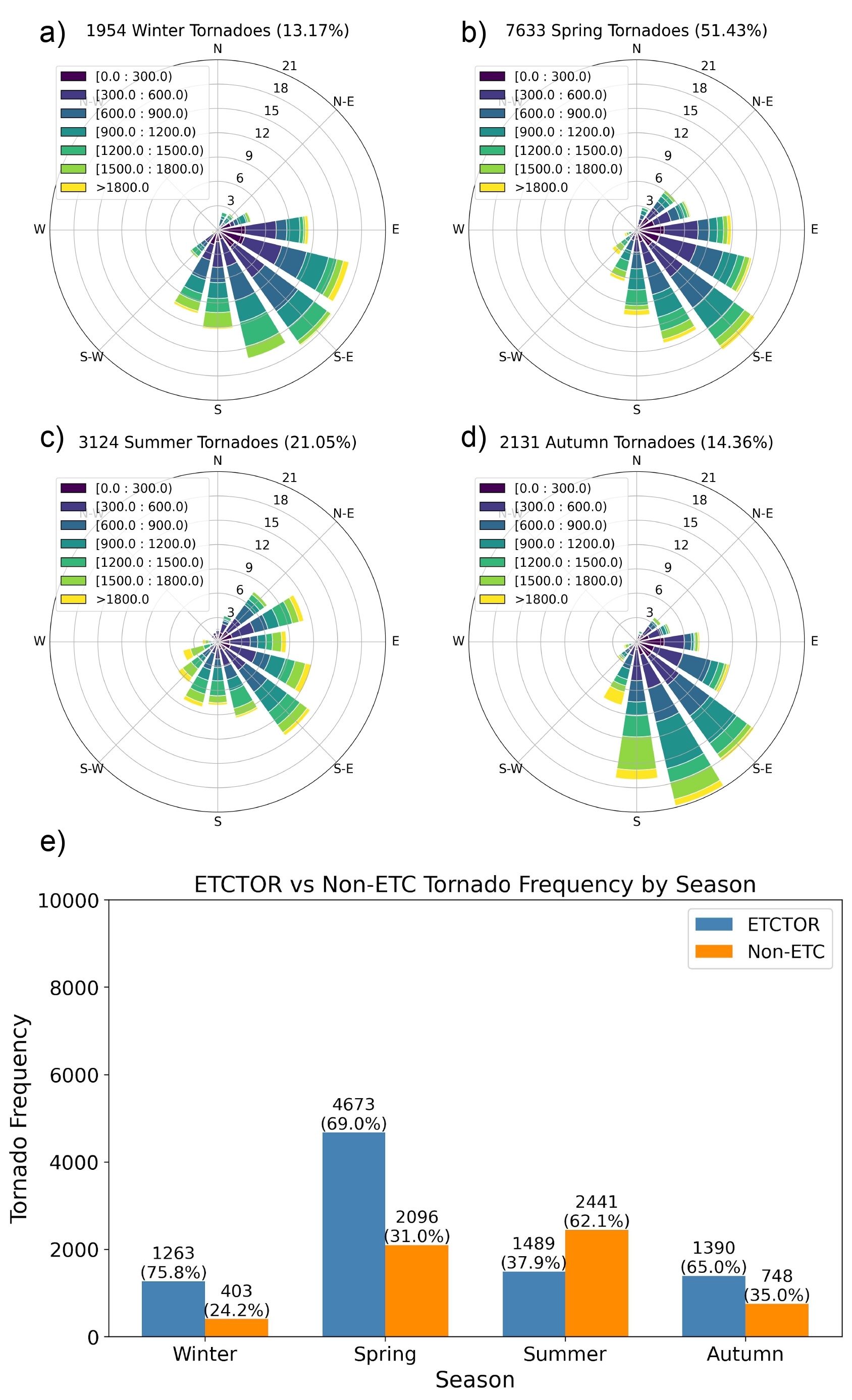}
    \caption{Distance and direction of ETCTOR relative to ETC track by season: (a) winter tornadoes; (b) spring tornadoes; (c) summer tornadoes; and (d) autumn tornadoes. (e) Frequency and percentage of ETCTOR vs. OUTTOR (defined as in Figure \ref{fig:assoc-2000}a) by season.}
    \label{fig:wind-season}
\end{figure}

Throughout these seasonal spatial shifts, ETCTOR generally form at a very consistent relative location due southeast of the ETC center (Figure \ref{fig:season-dist}a,b,c,d) and with similar ranges of variability. The range of distances between tornadoes and ETC centers is relatively constant, but there is a notable broadening of the range of directions relative to ETCs in summer relative to other seasons (Figure \ref{fig:wind-season}a-d), with an anomalously large fraction of tornadoes to the northeast of the ETC center as well as more often to the northwest of the low within 400 km. Autumn also exhibits a shift towards more tornadoes towards due south of the ETC center. Moreover, the percentage of ETCTOR versus non-ETC tornadoes (Figure \ref{fig:wind-season}e) varies systematically with season: largest in winter (75.8\%), smallest in summer (37.9\%) and comparable in spring and autumn (69\% and 65\%, respectively).

These two results indicate that tornadoes are most strongly associated with ETCs in winter and least strongly in summer. This outcome is consistent with the fact that the jet stream and ETCs are strongest in the winter when large-scale meridional temperature gradients are strongest and weakest in the summer when temperature gradients are weakest. The large-scale environment is also less conducive in the cooler months, such that in winter an ETC may be required to generate the environmental ingredients capable of producing tornadoes whereas in summer there are other pathways to generate these ingredients (e.g. a mesoscale convective system) in the absence of an ETC. Understanding how these tornadic environments are generated by ETCs and via other pathways is a valuable avenue of future work. Overall, seasonal variations in ETC-driven tornado activity are driven predominantly by seasonal variations in the ETC tracks themselves, especially outside of the summer season.

\subsection{Interannual variability and trends}\label{subsec:interannual}
\begin{figure}[t]
    \centering
    \includegraphics[width=1\linewidth]{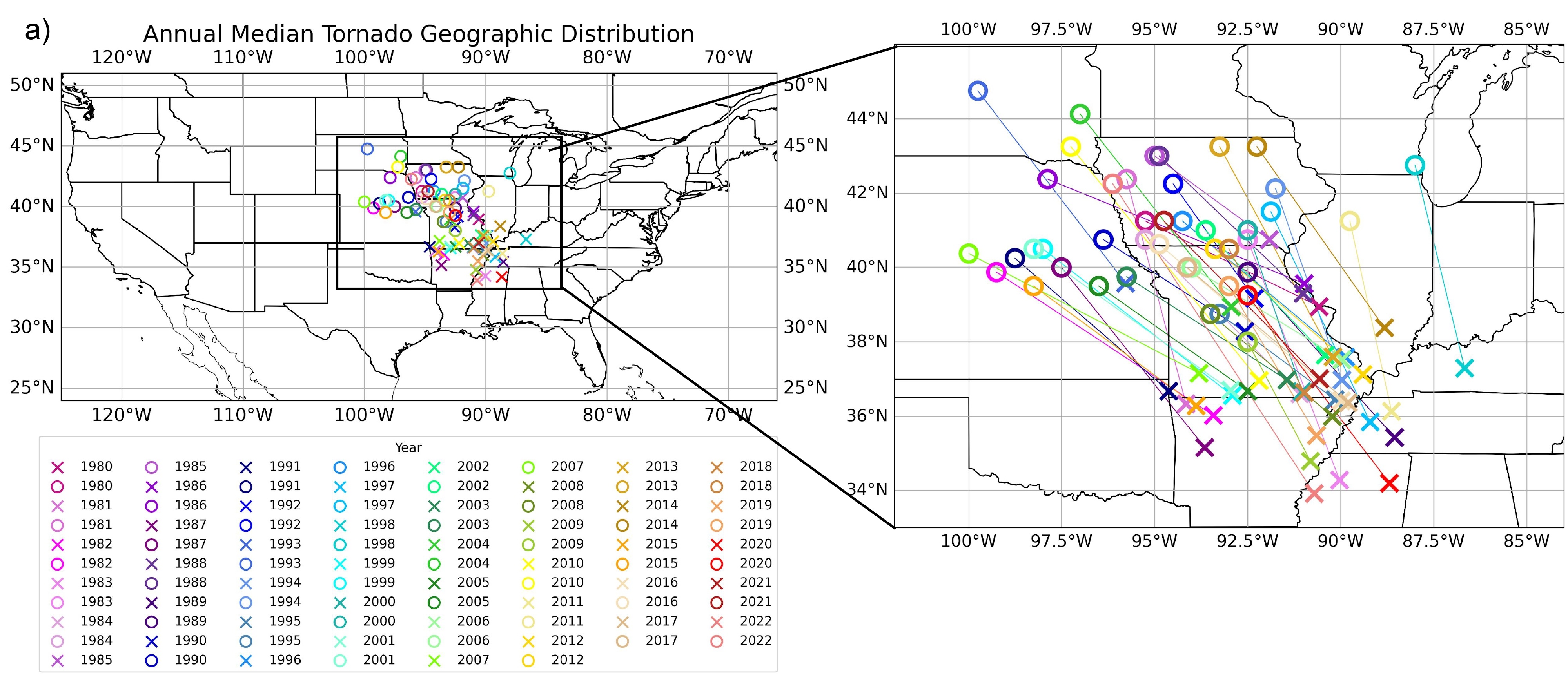}
    \noindent\includegraphics[width=1\linewidth]{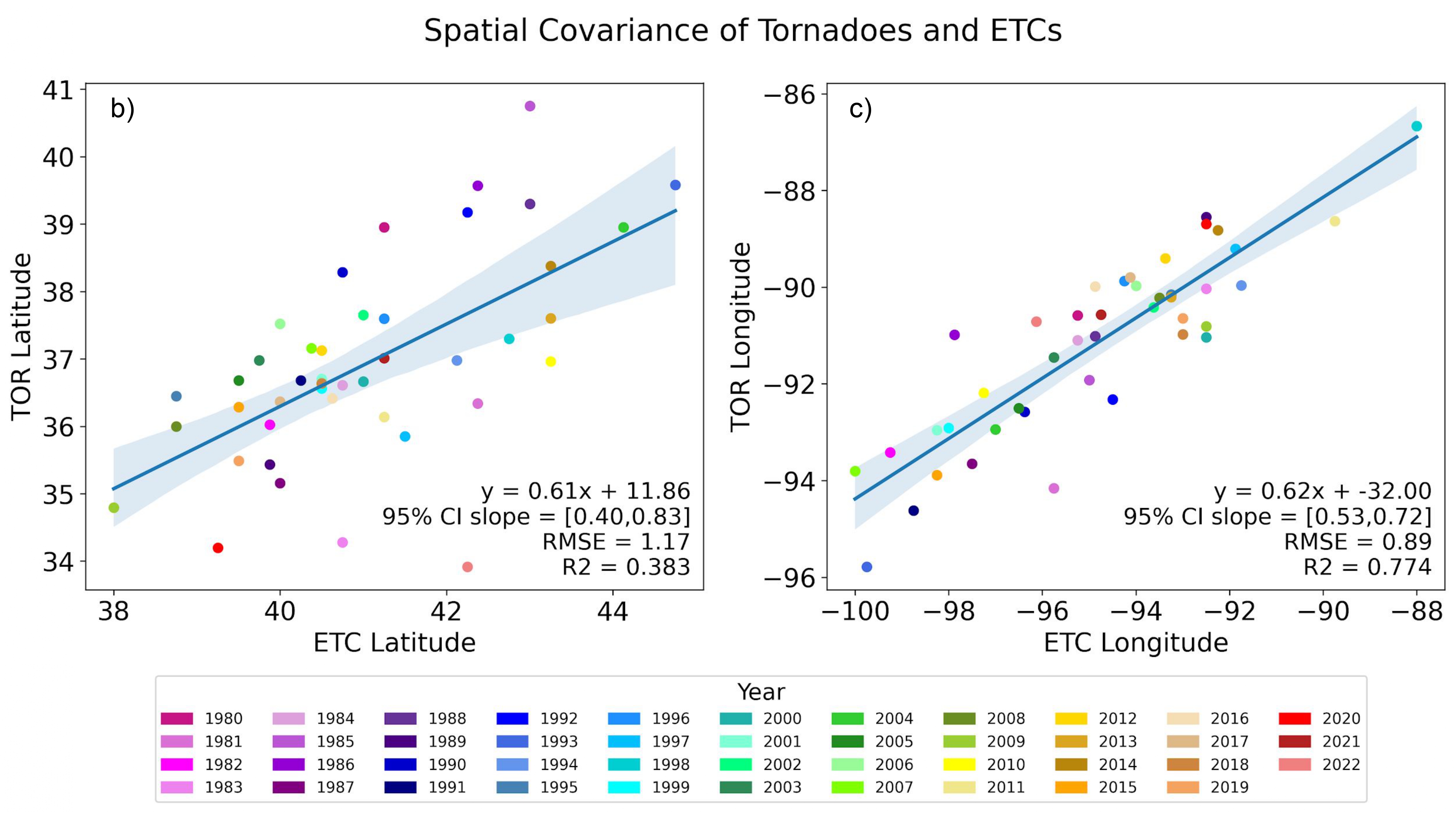}
    \caption{Interannual covariability in ETCTOR and associated ETCs. (a) Geographic distribution of the median latitude and longitude of tornadoes (colored Xs) and ETCs (colored circles) for each year during 1980-2022. (b) Scatterplot of median tornado latitude and median ETC latitude for each year with 95\% confidence bands (shading). (c) Same as (b) but for longitude.}
    \label{fig:annual-geo}
\end{figure}

We next examine the interannual variability in the relationship between ETCTOR and their associated ETCs during our period 1980-2022, with a focus on the median positions of each for simplicity. Throughout the period (Figure \ref{fig:annual-geo}a), median tornado and ETC locations vary within the domain (34N-45N, 85W-100W), where median tornado locations are centered over southeastern Missouri and median ETC locations are centered over southwestern Iowa. Hence, both vary over a similar range of approximately 7$^o$ latitude and 12$^o$ longitude, with the characteristic location of tornadoes to the southeast of the ETCs. Annual median tornado latitude and especially longitude both covary strongly and positively with associated ETC latitude (Figure \ref{fig:annual-geo}b) and longitude (Figure \ref{fig:annual-geo}c). Approximately 38\% of the variance in median tornado latitude is explained by median ETC latitude, while approximately 77\% in longitude.  Hence, the two indeed shift systematically with one another on interannual timescales, with joint east-west shifts being most pronounced. We performed this same analysis using all tornadoes, including those outside of our associated distance thresholds defined in Figure \ref{fig:assoc-2000}a, and found very similar results (not shown), increasing confidence in their covariability regardless of associated versus non-associated biases.


\begin{figure}[t]
    \centering
    \includegraphics[trim={1cm 0 1cm 0}, width=1\linewidth]{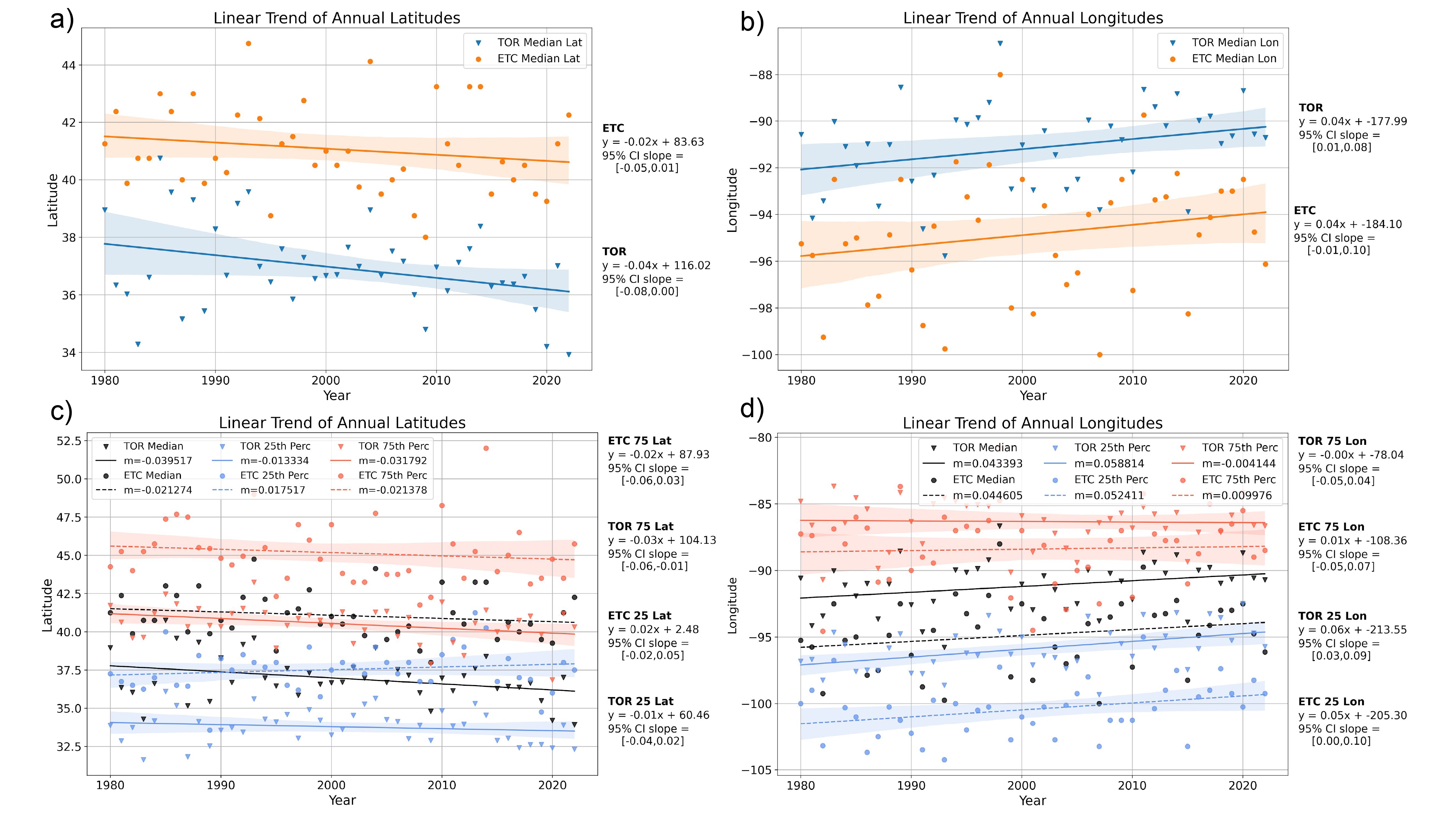}
    \noindent\includegraphics[width=1\linewidth]{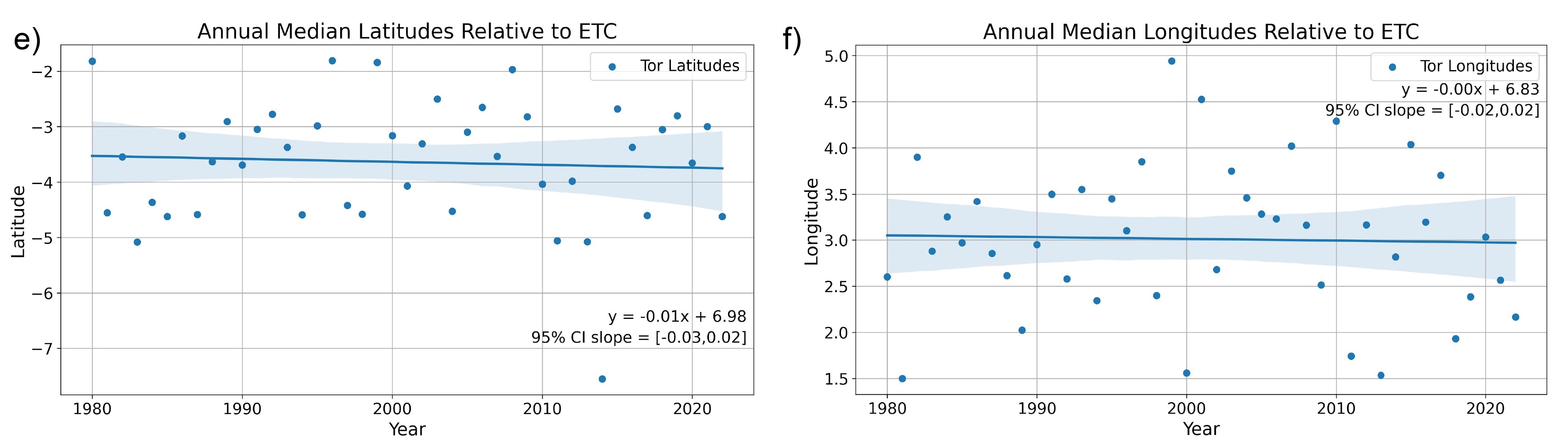}
    \caption{Long-term trends in annual median locations of ETCTOR and associated ETCs. (a) Annual latitude of tornadoes (blue) and associated ETCS (orange), with long-term trend (solid lines) and 95\% confidence bands (shading). (b) As in (a) but for longitude. (c) As in (a) but for the 25th and 75th percentiles of latitude (representing southern and northern range of variability, respectively) with median trend line from (a) overlaid in black for context. (d) As in (c) but for longitude (25th/75th representing western/eastern range of variability, respectively). (e) Median tornado latitude relative to ETC latitude. (f) As in (e) but for relative longitude. Linear regression results including 95\% confidence interval on the slopes included for each trend line.}
    \label{fig:trend}
\end{figure}

Finally, we examine long-term trends 1980-2022 in the location of ETCTOR and their associated ETCs (Figure \ref{fig:trend}) in the context of their strong interannual covariability and recent studies finding evidence of an eastward shift in tornadoes noted earlier. An open question is whether tornadoes could be shifting because associated ETCs are shifting or if there may be a shift in the trend in the location of tornadoes relative to the ETC centers (or both). Both the median tornado and median ETC locations show a statistically-significantly eastward trend with time and at similar rates (Figure \ref{fig:trend}b), both shifting eastward by 2 degrees (roughly 200 km) during the period. There is also a weaker southward trend in both the median ETC and the median tornado trends over time (Figure \ref{fig:trend}a) that is marginally significant. We further examine these trends by performing the same analysis on the 25th and 75th percentiles of each year (Figure \ref{fig:trend}c-d), where the former (values smaller than the median) represents the southward/westward ranges and the latter (larger than the median) represents the northward/eastward ranges. Figure \ref{fig:trend}d indicates that the 25th percentile is shifting steadily eastward whereas the 75th percentile is not for both tornadoes and ETC centers, indicating that the eastward shift is actually better interpreted as an eastward contraction of the zonal distribution, where there are simply fewer events to the west. Similarly, Figure \ref{fig:trend}c shows that the weak southward shift appears to be a southward meridional contraction as higher latitude tornado and associated ETC events are shifting southwards while lower latitude events are stable or even shifting northward in the ETC case, though these shifts are again only marginally significant. Finally, trends in the median location of tornadoes relative to associated ETC centers (Figure \ref{fig:trend}e-f) show no significant trend in either relative longitude or latitude, indicating that tornadoes are not shifting systematically closer to or farther from the ETC center.

Taken together, then, this analysis suggests that tornadoes are simply shifting eastwards along with their associated ETCs, driven primarily by a reduction in ETCs and tornadoes within the western half of the distribution over the Great Plains. Their close correspondences suggest that a shift in ETCs may be driving the shift in tornadoes. This is an important avenue for future analysis.



\subsection{ETC intensity}

\begin{figure}
    \centering
    \includegraphics[width=1\linewidth]{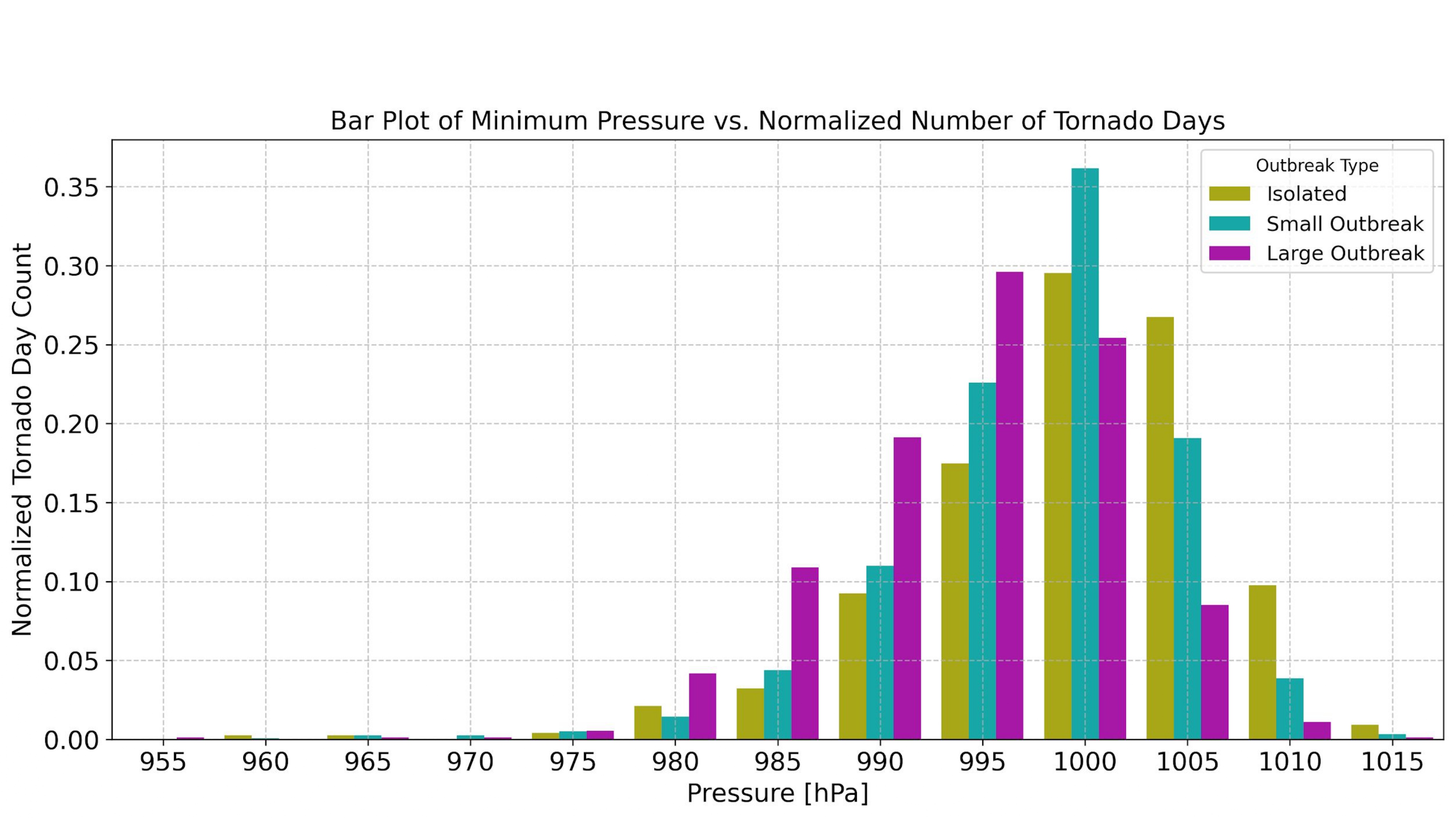}
    \caption{Bar plot showing ETC minimum pressure [hPa] (x-axis) and normalized number of tornado days (y-axis) for each outbreak size.}
    \label{fig:outbreak-pmin}
\end{figure}

Finally, we provide an initial analysis of how outbreak size depends on the intensity of the associated ETC as defined by its minimum central sea-level pressure. Figure \ref{fig:outbreak-pmin} displays the normalized distribution of tornado days as a function of minimum pressure of the associated ETC in 5 hPa bins for each outbreak size. While weaker ETCs with minimum pressures between 995 and 1005 hPa can be associated with tornado outbreaks of all sizes, there is a systematic shift in the distribution towards larger outbreaks for stronger ETCs (lower minimum pressures) over the range 980-1000 hPa. This shift is more subtle between isolated tornadoes and small outbreaks but is stronger between small and large outbreaks. For very intense ETCs below 980 hPa the sample size becomes too small to compare. This result suggests that larger tornado outbreaks tend to be associated with stronger cyclones. The topic of whether or not more intense ETCs are more capable of producing larger tornado outbreaks is worthy of much deeper study but lies beyond the scope of this analysis and so is left for future work.

\section{Conclusion}
This study analyzed the climatology of tornadoes in the United States relative to ETC using historical tornado data, ETC track data, and ERA5 reanalysis data for the period 1980-2022. We quantified the frequency of tornadoes associated with an ETC based on a subjective analysis of tornadoes that are likely to be in an ETC warm sector. We analyzed their spatial relationship and further examined how that relationship varies across outbreak size, tornado intensity, season, year, and ETC intensity. We found that approximately 72\% of tornadoes were associated with an ETC and most form broadly to the southeast and relatively close to the ETC center. Most ETCTOR form in large outbreaks of six or more tornadoes, and larger outbreaks tend to cluster more tightly together to the southeast of the ETC center. Spatial relationships do not vary with tornado intensity, likely because ETCs modulate tornado outbreaks and outbreaks with strong tornadoes produce many weaker ones too. Seasonal variability in tornado activity closely follow that of associated ETCs, both following the seasonal migration of the jet stream. Winter tornadoes have the highest percentage of ETCTOR when the jet stream and ETCs are strongest. ETCTOR also dominate in spring and autumn, but they are less common in summer when the jet stream and ETCs are weakest. Median tornado and associated ETC location covaries strongly on interannual timescales while tornado location relative to ETC center remains relatively constant. The eastward shift in the tornado distribution is associated with an eastward trend of comparable magnitude in associated ETCs while there is no trend in their relative locations, suggesting that recent trends in the spatial distribution of tornadoes may be driven by trends in associated ETCs. Finally, stronger ETCs are more capable of producing larger tornado outbreaks.

Overall, these results provide evidence that tornado activity is closely tied to ETC activity. Understanding where tornadoes typically occur in relationship to an extratropical cyclone can help with forecasting potential tornado occurrence. As the jet stream and ETCs change with a warming climate, we may be able to better predict how tornadoes may shift in the future. Changes in weather patterns change with climate variability across timescales may alter the regional patterns of tornado occurrence in ways that could be predictable if given knowledge of how ETC activity changes. Predicting tornado activity is especially important for vulnerable regions that may not be accustomed to dealing with such events.

There remain many viable avenues of future work. First and foremost, this work focused solely on the geographic relationships without tackling the underlying meteorology. Objective methods could be explored to define the warm sector itself through the ETC lifecycle to provide a more precise analysis of the spatial relationships within the warm sector and relative to warm/cold fronts or other synoptic features. Environmental parameters favorable for tornadoes could be calculated using ERA5 data to analyze the spatial structure of tornadic environments within ETC warm sectors, building on the work of \citet{tochimoto2022}. A deeper analysis of how our results and the associated meteorological conditions vary with ETC intensity and through the ETC lifecycle is a critical next step; we could only touch on the role of ETC intensity here. A detailed analysis of the role of the ETC in mediating the known linkage between jet-stream weather patterns and tornadoes, as well as the role of synoptic-scale features such as the Great Plains low-level jet, would be insightful. Exploration of those tornado cases that occur outside of an ETC warm sector would be of interest to understand alternative drivers. Finally, a reverse approach that includes all ETCs -- that both do and do not produce tornadoes -- and quantifies variations in ETC strength, trajectory, and the synoptic-scale climatology would also be helpful in understanding when, where, and how often ETCs produce tornado outbreaks.

\clearpage
\acknowledgments
This work was supported by NSF AGS grants 1945113 and 2209052. The authors thank Colin Zarzycki (PSU) for his assistance in updating the extratropical cyclone tracking code for this project, as well as Daniel J. Cecil for retrieving and providing the 1982-2007 tropical cyclone tornadoes database for the authors use. The authors are grateful to the National Center for Atmospheric Research (NCAR) Research Data Archive (RDA) for providing easy access to large reanalysis datasets, as well as to the Purdue Rosen Center for Advanced Computing (RCAC) for supercomputing resources and support.


%
%
\datastatement
Post-processed data and code for the figures in the manuscript will be made publicly available via the Purdue University Research Repository (PURR) prior to publication.

%






%



\bibliographystyle{ametsocV6}
\bibliography{references}

\end{document}